# Treatments of the exchange energy in density-functional theory

Tamás Gál

Section of Theoretical Physics, Institute of Nuclear Research of
the Hungarian Academy of Sciences,
4001 Debrecen, Hungary

**Abstract:** Following a recent work [Gál, Phys. Rev. A **64**, 062503 (2001)], a simple derivation of the density-functional correction of the Hartree-Fock equations, the Hartree-Fock-Kohn-Sham equations, is presented, completing an integrated view of quantum mechanical theories, in which the Kohn-Sham equations, the Hartree-Fock-Kohn-Sham equations and the ground-state Schrödinger equation formally stem from a common ground: density-functional theory, through its Euler equation for the ground-state density. Along similar lines, the Kohn-Sham formulation of the Hartree-Fock approach is also considered. Further, it is pointed out that the exchange energy of density-functional theory built from the Kohn-Sham orbitals can be given by degree-two homogeneous $N$-particle density functionals ($N$=1,2,...), forming a sequence of degree-two homogeneous exchange-energy density functionals, the first element of which is minus the classical Coulomb-repulsion energy functional.

Email address: galt@phys.unideb.hu

# I. Introduction

Density-functional theory (DFT) [1,2] forms the exact theoretical background for particle-density based approaches to the quantum mechanical many-body problem, such as theories of the Thomas-Fermi kind [3-7] and of the Slater kind [8-10]. The main idea behind DFT is to replace the quantum mechanical wave-function with the particle density as basic variable from which all properties of a quantum system can be gained, which means the use of density functionals instead of wave-function functionals. This idea is justified for ground states formally by the Hohenberg-Kohn theorems [11], but in practical applications DFT appears as an approximative theory because of the lack of knowledge of an exact expression for the ground-state energy density functional

$$E_v[\rho] = F[\rho] + \int \rho(\vec{r})v(\vec{r})d\vec{r} \quad , \qquad (1)$$

the minimum position of which gives the ground-state density for a given external potential $v(\vec{r})$ and particle number

$$N = \int \rho(\vec{r})d\vec{r} \quad , \qquad (2)$$

yielding the Euler equation

$$\frac{\delta E_v[\rho]}{\delta_N \rho(\vec{r})} = 0 \quad , \qquad (3a)$$

or

$$\frac{\delta F[\rho]}{\delta \rho(\vec{r})} + v(\vec{r}) = \mu \qquad (3b)$$

[with $\mu$ being the Lagrange multiplier corresponding to the $N$-conserving constraint on $\rho(\vec{r})$, Eq.(2)], for the determination of the ground-state density. In Eq.(1), $F[\rho]$, characteristic to the interaction between the (identical) Fermions of the considered system, is a universal functional of the density $\rho(\vec{r})$ [not depending on the external potential, $v(\vec{r})$], which represents the unknown in $E_v[\rho]$.

To treat a major part of $F[\rho]$ exactly, the Kohn-Sham (KS) method [12] introduces single-particle orbitals [ $u_i(\vec{r})$ ] into DFT, separating in $F$ the kinetic energy

$$T_s = \int \sum_{i=1}^{N} u_i^*(\vec{r}) \left(-\tfrac{1}{2}\nabla^2\right) u_i(\vec{r})\, d\vec{r} \tag{4}$$

of a ground-state system of noninteracting fermions with the density $\rho(\vec{r})$ of the given interacting fermion system, getting a set of single-particle equations,

$$-\tfrac{1}{2}\nabla^2 u_i(\vec{r}) + v_{KS}(\vec{r}) u_i(\vec{r}) = \varepsilon_i^{KS} u_i(\vec{r}) \qquad i=1,\ldots,N \tag{5}$$

with

$$v_{KS}(\vec{r}) = \frac{\delta(F[\rho] - T_s[\rho])}{\delta\rho(\vec{r})} + v(\vec{r})\ , \tag{6}$$

for the determination of the (ground-state) density

$$\rho(\vec{r}) = \sum_{i=1}^{N} u_i^*(\vec{r})\, u_i(\vec{r})\ , \tag{7}$$

instead of using Eq.(3) directly. With the subtraction of the classical Coulomb repulsion

$$J[\rho] = \frac{1}{2}\iint \frac{\rho(\vec{r})\rho(\vec{r}\,')}{|\vec{r} - \vec{r}\,'|} d\vec{r}\, d\vec{r}\,' \tag{8}$$

from $F[\rho] - T_s[\rho]$,

$$F[\rho] = T_s[\rho] + J[\rho] + E_{xc}[\rho]\ , \tag{9}$$

then only the relatively small exchange-correlation (xc) part of the energy remains to be approximated. The exchange part of $E_{xc}$ is usually defined by

$$E_x = -\frac{1}{2}\iint \sum_{i,j=1}^{N} \frac{u_i^*(\vec{r})u_j(\vec{r})u_j^*(\vec{r}\,')u_i(\vec{r}\,')}{|\vec{r} - \vec{r}\,'|} d\vec{r}\, d\vec{r}\,'\ . \tag{10}$$

A further reduction of need for approximation can be achieved by utilizing the definition Eq.(10) of the exchange energy by single-particle orbitals, and setting up the single-particle equations

$$-\frac{1}{2}\nabla^2 u_i(\vec{r}) - \int \sum_{j=1}^{N} \frac{u_j(\vec{r}) u_j^*(\vec{r}\,') u_i(\vec{r}\,')}{|\vec{r} - \vec{r}\,'|} d\vec{r}\,' + v_{HFKS}(\vec{r}) u_i(\vec{r}) = \varepsilon_i^{HFKS} u_i(\vec{r}) \qquad i=1,\ldots,N\ , \tag{11}$$

with the local (multiplicative) potential

$$v_{HFKS}(\vec{r}) = \frac{\delta(F[\rho] - T_s^{HFKS}[\rho] - E_x^{HFKS}[\rho])}{\delta\rho(\vec{r})} + v(\vec{r}) = \frac{\delta J[\rho]}{\delta\rho(\vec{r})} + \frac{\delta E_c^{HFKS}[\rho]}{\delta\rho(\vec{r})} + v(\vec{r})\ , \tag{12}$$

which gives a density-functional correction of the Hartree-Fock (HF) equations, yielding the Hartree-Fock-Kohn-Sham method [12,13]. The solutions $\{u_i(\vec{r})\}_{i=1}^{N}$ of the different sets of single-particle equations, Eqs.(5) and (11), corresponding to the same $\rho(\vec{r})$, are different. Note that though with the HFKS approach a higher degree of exactness of practical calculations can be reached, it means another step of backing away from a pure *density-functional* theory; in addition to the introduction of orbitals, now a nonlocal potential appearing. (The next step would be to "introduce" many-particle wave-functions to treat correlation exactly as well, regaining the Schrödinger equation itself.)

Following the spirit of a recent work by the author [14], in this article a simple derivation of the HFKS equations will be presented, placing them in a unified scheme of quantum mechanical theories in which the Kohn-Sham equations, the Hartree-Fock-Kohn-Sham equations and the ground-state Schrödinger equation originate from one root: the Hohenberg-Kohn theorems, through the Euler equation Eq.(3). Beside the HFKS scheme, another mixture of DFT and the Hartree-Fock theory, the Kohn-Sham formulation of the Hartree-Fock scheme, will also be considered. In the second part of the paper, the concept of *N*-particle exchange-energy density functionals will be introduced, following also [14] (see also [15]), and it will be shown that they emerge naturally as second-degree homogeneous functionals, for every particle number *N*. For simplicity, spin is not considered throughout the paper.

## II. The Hartree-Fock-Kohn-Sham scheme, and the Kohn-Sham formulation of the Hartree-Fock theory

Building on the idea of the constrained search definition of density functionals [16-18], the exact incorporation of correlation into the Hartree-Fock method in the form of a multiplicative, density-dependent potential (which is a density-functional derivative), that is, the so-called Hartree-Fock-Kohn-Sham method, can be derived directly from density-functional theory in a simple way.

The starting point is the definition of the functional [17]

$$F_D[\rho] = \min_{\psi_D \to \rho} \{T[\psi_D] + V[\psi_D]\} \;, \tag{13}$$

which, for a $\rho(\vec{r})$, gives the minimum of the sum $T[\psi]+V[\psi]$ of wave-function functionals

$$T[\psi] = \left\langle \psi \left| \sum_{i=1}^{N} -\tfrac{1}{2}\nabla_i^2 \right| \psi \right\rangle \tag{14}$$

and

$$V[\psi] = \left\langle \psi \left| \sum_{i<j}^{N} \frac{1}{|\vec{r}_i - \vec{r}_j|} \right| \psi \right\rangle \tag{15}$$

over the set of normalized Slater determinants ($\psi_D$) that yield $\rho(\vec{r})$, that is, $\left\langle \psi_D | \hat{\rho}(\vec{r}) | \psi_D \right\rangle = \rho(\vec{r})$ —while $F[\rho]$, defined by

$$F[\rho] = \min_{\psi \to \rho} \{T[\psi] + V[\psi]\} \;, \tag{16}$$

searches over the space of normalized (antisymmetric) wave-functions not restricted to be determinants. Similar to the case of the noninteracting kinetic-energy density functional

$$T_s[\rho] = \min_{\psi_D \to \rho} T[\psi_D] \;, \tag{17}$$

Eq.(13), by the mapping

$$\rho \to \psi_D : \quad \rho[\psi_D] = \rho, \quad T[\psi_D] + V[\psi_D] = F_D[\rho] \;, \tag{18}$$

associates every $\rho(\vec{r})$ with a fictive noninteracting system of density $\rho(\vec{r})$, composed of single-particle orbitals $u_i(\vec{r})$ of number $N = \int \rho(\vec{r})d\vec{r}$. Writing Eq.(13) explicitly with the orbitals,

$$F_D[\rho] = \min_{\{u_i\} \to \rho} \left\{ T_s[u_1, u_1^*, ..., u_N, u_N^*] + E_x[u_1, u_1^*, ..., u_N, u_N^*] \right\} + J[\rho] \;, \tag{19}$$

with $T_s[u_1, u_1^*, ..., u_N, u_N^*]$ being defined by Eq.(4) and $E_x[u_1, u_1^*, ..., u_N, u_N^*]$ by Eq.(10). The orbitals $u_i(\vec{r})$ corresponding to a density $\rho(\vec{r})$ can be determined through an Euler-Lagrange minimization procedure [16,19,20] (see also p. 151 and p. 187 of [1]), with the constraints of fixed density, and orthonormalization,

$$\int u_i^*(\vec{r}) \, u_j(\vec{r}) d\vec{r} = \delta_{ij} \;, \tag{20}$$

yielding the Euler-Lagrange equations

$$-\frac{1}{2}\nabla^2 u_i(\vec{r}) - \int \sum_{j=1}^{N} \frac{u_j(\vec{r})u_j^*(\vec{r}\,')}{|\vec{r} - \vec{r}\,'|} u_i(\vec{r}\,')d\vec{r}\,' + \lambda_\rho(\vec{r})u_i(\vec{r}) = \varepsilon_i u_i(\vec{r}) \qquad i=1,...,N \tag{21}$$

(in canonical form), with the Lagrange multipliers $\varepsilon_i$ securing the normalization of $u_i(\vec{r})$ and $\lambda_\rho(\vec{r})$ securing the fulfilment of Eq.(7) (with $\rho(\vec{r})$ fixed). This approach of the minimization problem posed by Eq.(19), however, tells nothing more about $\lambda_\rho(\vec{r})$, about its connection to $\rho(\vec{r})$, and hence to $v_{HFKS}(\vec{r})$; this is because, as has been pointed out in [14], it does not utilize the definition Eq.(19), the implicit definition of the $u_i(\vec{r})$'s corresponding to a given $\rho(\vec{r})$, fully.

With a different approach, namely, minimizing the difference

$$\Delta_{F_D}[u_1, u_1^*, \ldots, u_N, u_N^*] := T_s[u_1, u_1^*, \ldots, u_N, u_N^*] + E_x[u_1, u_1^*, \ldots, u_N, u_N^*]$$
$$- (F_D - J)[\rho[u_1, u_1^*, \ldots, u_N, u_N^*]] \quad (22)$$

under the orthonormalization constraint only (instead of the above minimization of $T_s[u_1, u_1^*, \ldots, u_N, u_N^*] + E_x[u_1, u_1^*, \ldots, u_N, u_N^*]$), the $\rho$-dependence of the multiplier that forces the orbitals $u_i(\vec{r})$ to yield a given $\rho(\vec{r})$ can be identified: the Euler-Lagrange equations resulting from the minimization of $\Delta_{F_D}[u_1, u_1^*, \ldots, u_N, u_N^*]$, under normalization constraint on $u_i(\vec{r})$'s, are

$$-\frac{1}{2}\nabla^2 u_i(\vec{r}) - \int \sum_{j=1}^{N} \frac{u_j(\vec{r}) u_j^*(\vec{r}')}{|\vec{r}-\vec{r}'|} u_i(\vec{r}') d\vec{r}' - \frac{\delta(F_D[\rho]-J[\rho])}{\delta\rho(\vec{r})} u_i(\vec{r}) = \varepsilon_i u_i(\vec{r}) \quad i=1,\ldots,N. \quad (23)$$

The basis for this minimization is that, following from the definition of $F_D[\rho]$,

$$\Delta_{F_D}[u_1, u_1^*, \ldots, u_N, u_N^*] \geq 0 \quad (24)$$

for any normalized $u_i(\vec{r})$'s, the equality holding for the orbitals $\{u_i(\vec{r})\}_{i=1}^{N}$ associated to $\rho(\vec{r})$ by Eq.(18); thus only the normalization of $u_i(\vec{r})$'s has to be reckoned with as external constraint. With Eq.(23) then the equations Eq.(11) of the HFKS method arise immediatelly through the Hohenberg-Kohn Euler equation, Eq.(3), that is,

$$\frac{\delta(F_D[\rho]-J[\rho])}{\delta\rho(\vec{r})} + v_{HFKS}(\vec{r}) = \mu \quad (25)$$

with

$$v_{HFKS}(\vec{r}) = \frac{\delta\{F[\rho]-(F_D[\rho]-J[\rho])\}}{\delta\rho(\vec{r})} + v(\vec{r}) \quad (26)$$

[Eq.(12)], which brings the external potential $v(\vec{r})$ into the equations, fixing $\rho(\vec{r})$ (as that corresponding to $v(\vec{r})$).

It is important to underline that the correlation energy functional of the HFKS scheme is not the same as that of the Kohn-Sham theory, since it is defined by

$$E_c^{HFKS}[\rho] = F[\rho] - F_D[\rho] \,, \tag{27}$$

while in the Kohn-Sham scheme

$$E_c[\rho] = F[\rho] - T_s[\rho] - J[\rho] - E_x[\rho] \,, \tag{28}$$

but

$$F_D[\rho] \neq T_s[\rho] + J[\rho] + E_x[\rho] \tag{29}$$

[$E_x[\rho]$ is defined through Eq.(10) with $u_i(\vec{r})$'s being the ($\rho$-dependent) orbitals of the Kohn-Sham scheme]. In particular, since $F_D[\rho]$ minimizes $T_s[u_1, u_1^*, ..., u_N, u_N^*] + E_x[u_1, u_1^*, ..., u_N, u_N^*]$ in total [under Eq.(7)],

$$F_D[\rho] \leq T_s[\rho] + J[\rho] + E_x[\rho] \,, \tag{30}$$

from which

$$E_c^{HFKS}[\rho] \geq E_c[\rho] \,, \tag{31}$$

which means that, besides the need for approximation of $E_x[\rho]$, in the Kohn-Sham method, there is also a greater correlation part in the energy density functional to be approximated. (Note that both kinds of correlation energies are always negative or zero [13].)

Beside the density-functional amplification of Hartree-Fock theory, discussed so far, there is another mixture of DFT and Hartree-Fock theory: the Kohn-Sham formulation of the Hartree-Fock scheme [19,17,21,13] (the Kohn-Sham-Hartree-Fock scheme), where the nonlocal exchange operator of the Hartree-Fock equations is replaced by a local potential which is a density-functional derivative. This theory gives the sound justification of early attempts in that direction. As the Hartree-Fock method is an approximative theory by definition, neglecting correlation completely by narrowing the space of wave-functions to Slater determinants, its Kohn-Sham formulation is not exact either. Here, the density functional $F[\rho]$ of exact DFT is replaced by $F_D[\rho]$, and the energy functional

$$E_v^{HF}[\rho] = F_D[\rho] + \int \rho(\vec{r}) v(\vec{r}) d\vec{r} \tag{32}$$

is minimized, under the *N*-conservation constraint, giving the Hartree-Fock ground-state energy, and density. This yields the Euler equation

$$\frac{\delta F_D[\rho]}{\delta \rho(\vec{r})} + v(\vec{r}) = \mu^{HF} \qquad (33)$$

for the determination of the Hartree-Fock ground-state density. Note that, with $E_v^{HF}[\rho]$, $E_v[\rho]$ arises as

$$E_v[\rho] = E_v^{HF}[\rho] + E_c^{HFKS}[\rho] \ . \qquad (34)$$

Separating $T_s[\rho]$ in $E_v^{HF}[\rho]$, analogously to exact DFT, a major part of the unknown $F_D[\rho]$ can be treated exactly, obtaining the single-particle Schrödinger equations

$$-\tfrac{1}{2}\nabla^2 u_i(\vec{r}) + v_{KS}^{HF}(\vec{r}) u_i(\vec{r}) = \varepsilon_i^{KSHF} u_i(\vec{r}) \qquad i=1,\ldots,N, \qquad (35)$$

with Eq.(7), for the determination of the density, where

$$v_{KS}^{HF}(\vec{r}) = \frac{\delta(F_D[\rho] - T_s[\rho])}{\delta \rho(\vec{r})} + v(\vec{r}) \ . \qquad (36)$$

These equations can be obtained in a simple manner, utilizing the idea described above in connection with the HFKS method; namely, minimizing the difference

$$\Delta_{T_s}[u_1, u_1^*, \ldots, u_N, u_N^*] := T_s[u_1, u_1^*, \ldots, u_N, u_N^*] - T_s[\rho[u_1, u_1^*, \ldots, u_N, u_N^*]] \qquad (37)$$

under orthonormalization constraint on $u_i(\vec{r})$'s (just as in the case of the derivation of the Kohn-Sham equations given in [14]), then using the pure density-functional Euler equation Eq.(33) to substitute for $-\frac{\delta T_s[\rho]}{\delta \rho(\vec{r})}$ in the obtained Euler-Lagrange equations, hereby fixing $\rho(\vec{r})$ by the external potential. The traditional quantum mechanical equations of the Hartree-Fock approach, the Hartree-Fock equations themselves, can be recovered as well from Eq.(33), following the above procedure, but now

$$\Delta_{F_D}[u_1, u_1^*, \ldots, u_N, u_N^*] = T_s[u_1, u_1^*, \ldots, u_N, u_N^*] + E_x[u_1, u_1^*, \ldots, u_N, u_N^*] - (F_D - J)[\rho[u_1, u_1^*, \ldots, u_N, u_N^*]] \qquad (38)$$

has to be minimized (followed by the insertion of Eq.(33)); just as can the (ground-state) Schrödinger equation of exact quantum mechanics be derived from the Hohenberg-Kohn Euler equation [Eq.(3)] [14]. Note that the difference functionals Eq.(22) and Eq.(38) are the same; the difference between the single-particle equations of the HFKS and the HF theories arises from the different [exact, or approximate (HF)] physics behind the equations, that is, different DFT backgrounds, characterized by the density-functional Euler equations Eq.(3) and Eq.(33), respectively.

Regarding the derivations of Schrödinger equations from DFT, some words need to be said about the density-functional derivatives (local potentials) appearing in these equations since the density functionals defined through wave-functions are defined only for densities of integer norm $N$. This problem can be eliminated with the help of the concept of $N$-particle density functionals and $N$-conserving functional differentiation [22] (for a discussion of its mathematics, see [23]). In [14], it has been pointed out that the Euler-Lagrange minimization procedure proposed there, and used above as well, to determine the corresponding orbitals for a density $\rho(\vec{r})$, can be carried out by using an $N$-particle density functional $A_N[\rho]$ instead of the given density functional $A[\rho]$, defining the difference functional $\Delta$ with $A_N[\rho]$, $\Delta_{A_N}[u_1,u_1^*,...,u_N,u_N^*] := A[u_1,u_1^*,...,u_N,u_N^*] - A_N[\rho[u_1,u_1^*,...,u_N,u_N^*]]$, since the normalization of the orbitals and the fixation of their number conserve the norm $N$ of $\rho(\vec{r})$. An $A_N[\rho]$ defined for every $\rho(\vec{r})$ can then be chosen; the trivial constant shifting [22] (or other homogeneous extension) of $A[\rho_N]$, e.g., is one proper choice, being (unconstrained) differentiable if $A[\rho]$ is differentiable over the space of $\rho(\vec{r})$'s of the given norm $N$. In the orbital equations resulting from the minimization of $\Delta_{A_N}[u_1,u_1^*,...,u_N,u_N^*]$, adding $\frac{1}{N}\int \rho(\vec{r}\,') \frac{\delta A_N[\rho]}{\delta \rho(\vec{r}\,')} d\vec{r}\,' u_i(\vec{r})$ to both sides of the equations [22], the unconstrained derivative of $A_N[\rho]$ is replaced by the $N$-conserving functional derivative, $\frac{\delta A_N[\rho]}{\delta_N \rho(\vec{r})}$, for which [22]

$$\frac{\delta A_N[\rho_N]}{\delta_N \rho(\vec{r})} = \frac{\delta A[\rho_N]}{\delta_N \rho(\vec{r})} \quad . \tag{39}$$

After using Eq.(39), finally, the replacement of the unconstrained derivative by the corresponding $N$-conserving derivative (of $A[\rho]$) is achieved in the orbital equations (for $\rho_N(\vec{r})$'s), thereby restricting the differentiation to the allowed domain of $\rho(\vec{r})$'s. Note that similar considerations hold also if $A[\rho]$ comes from a wave-function functional $A[\psi]$ that is not an orbital functional. With $N$-conserving density-functional derivatives, the derivations of Schrödinger equations from DFT thus look like: from

$$-\frac{1}{2}\nabla^2 u_i(\vec{r}) - \int \sum_{j=1}^{N} \frac{u_j(\vec{r})u_j^*(\vec{r}\,')}{|\vec{r}-\vec{r}\,'|} u_i(\vec{r}\,')d\vec{r}\,' - \frac{\delta(F_D[\rho_N]-J[\rho_N])}{\delta_N \rho(\vec{r})} u_i(\vec{r}) = \varepsilon_i' u_i(\vec{r}) \qquad (40)$$

(e.g.), using

$$\frac{\delta F[\rho]}{\delta_N \rho(\vec{r})} + v(\vec{r}) = \mu' \;, \qquad (41)$$

$$-\frac{1}{2}\nabla^2 u_i(\vec{r}) - \int \sum_{j=1}^{N} \frac{u_j(\vec{r})u_j^*(\vec{r}\,')}{|\vec{r}-\vec{r}\,'|} u_i(\vec{r}\,')d\vec{r}\,' + \left\{\frac{\delta J[\rho_N]}{\delta_N \rho(\vec{r})} + \frac{\delta E_c^{HFKS}[\rho_N]}{\delta_N \rho(\vec{r})} + v(\vec{r})\right\} u_i(\vec{r}) = \varepsilon_i'^{HFKS} u_i(\vec{r}) \;. \quad (42)$$

Another way of treating the problem of $A[\rho]$ ($= T_s[\rho]$, $F_D[\rho]$, or $F[\rho]$) being defined only for integer $N$'s is to utilize that a chain rule, namely,

$$\frac{\delta A[\rho[g]]}{\delta g(x)} = \int \left.\frac{\delta A[\rho]}{\delta \rho(x')}\right|_K \frac{\delta \rho(x')}{\delta g(x)} dx' \;,$$

embracing

$$\frac{\delta A[\rho[g]]}{\delta g(x)} = \int \frac{\delta A[\rho]}{\delta_K \rho(x')} \frac{\delta \rho(x')}{\delta g(x)} dx' \;,$$

still holds (as proved in the Appendix of [23]) for cases $A[\rho[g]]$ where the functional $\rho[g]$ is such that $\rho(x)[g]$ satisfies the given $K$-conservation constraint –here $\int \rho(\vec{r})d\vec{r} = N$– for any (allowed) $g(x')$. In the present cases, $g=(u_1,u_1^*,...,u_N,u_N^*)$ [or $g=(\psi,\psi^*)$], and the normalization of $u_i(\vec{r})$'s ensures the fulfilment of $\int \rho(\vec{r})d\vec{r} = N$. Thus,

$$\left.\frac{\delta F_D[\rho_N[u_j,u_j^*]]}{\delta u_i(\vec{r})}\right|_{\int u_i^*(\vec{r})u_i(\vec{r})d\vec{r}=1} = \int \left.\frac{\delta F_D[\rho_N]}{\delta \rho(\vec{r}\,')}\right|_N \left.\frac{\delta \rho(\vec{r}\,')}{\delta u_i(\vec{r})}\right|_{\int u_i^*(\vec{r})u_i(\vec{r})d\vec{r}=1} d\vec{r}\,' \;, \qquad (43a)$$

or

$$\left.\frac{\delta F_D[\rho_N[u_j,u_j^*]]}{\delta u_i(\vec{r})}\right|_{\int u_i^*(\vec{r})u_i(\vec{r})d\vec{r}=1} = \int \frac{\delta F_D[\rho_N]}{\delta_N \rho(\vec{r}\,')} \left.\frac{\delta \rho(\vec{r}\,')}{\delta u_i(\vec{r})}\right|_{\int u_i^*(\vec{r})u_i(\vec{r})d\vec{r}=1} d\vec{r}\,' \;. \qquad (43b)$$

Note that this chain rule is essential also for the variational derivation of the Kohn-Sham equations themselves.

It is important to note here finally that, having the *N*-conserving derivatives (as potentials) in the derived wave-function equations, the fractional-particle-number generalization [24] of density functionals can be substituted for the original density functionals, obtaining the possibility of relaxing the constraint of *N*-conservation on the functional differentiation and using the unconstrained (left or right) derivatives of density functionals.

## III. Second-degree homogeneous *N*-particle exchange-energy density functionals

To construct approximations for density functionals, knowledge about the structure of the functionals is essential. In this section, a natural formal construction of the exchange-energy density functional $E_x[\rho]$ of Kohn-Sham DFT, via second-degree homogeneous *N*-particle exchange-energy density functionals, will be proposed.

It is known that $E_x[\rho]$, defined through Eq.(10) with $u_i(\vec{r})$ as the ($\rho$-dependent) Kohn-Sham orbitals, scales homogeneously of degree one with coordinate scaling [25], that is,

$$E_x[\lambda^3 \rho(\lambda \vec{r})] = \lambda \, E_x[\rho(\vec{r})] \;, \tag{44}$$

from which

$$E_x[\rho] = -\int \rho(\vec{r}) \vec{r} \nabla \frac{\delta E_x[\rho]}{\delta \rho(\vec{r})} d\vec{r} \tag{45}$$

emerges [26]. Beside this property, the question of homogeneity of $E_x[\rho]$ in $\rho$ itself may also arise, considering that the classical Coulomb repulsion functional $J[\rho]$, the negative of which is an exact expression for $E_x$ for one-particle densities, is homogeneous of degree two, i.e.

$$J[\lambda \rho] = \lambda^2 J[\rho] \;. \tag{46}$$

The question of homogeneity with respect to the density, and related, or affinitative, properties of the various components of the energy, has been addressed in several works recently [27-39], homogeneity being a very attractive (strong) analytical property for functionals. For the noninteracting kinetic-energy density functional $T_s[\rho]$, the first-degree homogeneity in $\rho$ [33] has been shown to be an incorrect proposal [37,38], though the Weizsäcker (one-particle kinetic-energy) functional,

$$T_W[\rho] = \frac{1}{8}\int \frac{|\nabla\rho(\vec{r})|^2}{\rho(\vec{r})}d\vec{r} \tag{47}$$

(which is an exact expression for $T_s[\rho]$ for one-particle densities), is homogeneous of degree one (in $\rho$). In the case of $E_x[\rho]$, the invalidity of homogeneity (of second, or any other, degree) in $\rho$ can be proven in a simple way, as for $T_s[\rho]$ [37]; since if homogeneity of a degree $k$ held then

$$\left(\frac{1}{N}\right)^k E_x[\rho] = E_x\left[\frac{\rho}{N}\right] = -J\left[\frac{\rho}{N}\right] = -\left(\frac{1}{N}\right)^2 J[\rho] , \tag{48}$$

that is, $E_x[\rho]$ would be $-N^{k-2}J[\rho]$. Note that from the second-degree homogeneity of $E_x[\rho]$ (in $\rho$), Eq.(15) of [39], derived from the result of [32] for the full electron-electron-interaction energy density functional, would also follow. (Chan and Handy [37] having shown that the homogeneity results of [32-34] are incorrect in the case of the kinetic energy, Joubert [39] has also shown that for the correlation energy functional after, in [38], it was pointed out that an essential question was not treated properly in [32-34], therefore their results cannot be considered to be valid.)

With the introduction of the concept of *N*-particle density functionals, that is, functionals that are valid expressions for a given density functional for *N*-particle densities, however, it can be shown that the exchange energy can be given by density functionals homogeneous of degree two, as in the case of $T_s$, which can be given by first-degree homogeneous *N*-particle density functionals, for all particle numbers *N* [14]. By utilizing the result in [14] that the KS orbitals arise most naturally as degree-$\frac{1}{2}$ homogeneous functionals of the density, that is,

$$\bigl(u_1(\vec{r}),...,u_N(\vec{r})\bigr)[k\rho] = \sqrt{k}\bigl(u_1(\vec{r}),...,u_N(\vec{r})\bigr)[\rho] , \tag{49}$$

$E_x$ arises as a second-degree homogeneous functional of the density, through Eq.(10). The resultant density functional is an *N*-particle density functional, $E_x^N[\rho]$, since its form depends on the number *N* of orbitals used in its construction. This $E_x^N[\rho]$ thus satisfies

$$E_x^N[\lambda\rho] = \lambda^2 E_x^N[\rho] \ , \tag{50}$$

that is,

$$E_x^N[\rho] = \frac{1}{2}\int \rho(\vec{r}) \frac{\delta E_x^N[\rho]}{\delta\rho(\vec{r})} d\vec{r} \ . \tag{51}$$

Note that an *N*-particle exchange-energy density functional (as any other *N*-particle density functional) can be considered as a two-variable functional, $E_x[N,\rho]$, and the general exchange-energy density functional emerges as

$$E_x[\rho] = E_x[\int \rho, \rho] \ , \tag{52}$$

for which thus the homogeneity Eq.(50) yields the property

$$E_x[\rho] = \frac{1}{\lambda^2} E_x[\int \rho, \lambda\rho] \ . \tag{53}$$

Beside Eq.(51), also the analogue of Eq.(45) can be written for $E_x^N[\rho]$, for *N*-particle densities [$\rho_N(\vec{r})$],

$$E_x^N[\rho_N] = -\int \rho_N(\vec{r})\vec{r}\nabla \frac{\delta E_x^N[\rho_N]}{\delta\rho(\vec{r})} d\vec{r} \ , \tag{54}$$

since the difference between the derivatives $\frac{\delta E_x[\rho_N]}{\delta\rho(\vec{r})}$ and $\frac{\delta E_x^N[\rho_N]}{\delta\rho(\vec{r})}$ is $\vec{r}$-independent [22,23]. Eq.(54) is in accordance with the fact that coordinate scaling conserves the normalization of the density, with the use of which

$$E_x^N[\lambda^3\rho_N(\lambda\vec{r})] = E_x[\lambda^3\rho_N(\lambda\vec{r})] = \lambda E_x[\rho_N(\vec{r})] = \lambda E_x^N[\rho_N(\vec{r})] \tag{55}$$

[giving also Eq.(54)]; see the Appendix for the proof for a general $\rho(\vec{r})$.

It is worth noting that requiring *N*-particle density functionals to be homogeneous of some degree eliminates the ambiguity present in their definition [15], meaning that the only possible *N*-particle noninteracting kinetic-energy and exchange-energy density functionals homogeneous of degree one and two, respectively, are those constructed in [14] and here (see also the Appendix).

The simplest exchange energy density functional arising from the conditions of homogeneity of degree two in density scaling [Eq.(50)] and homogeneity of degree one in coordinate scaling [Eq.(55), see also Eq.(A6)] is minus the Coulomb functional $J[\rho]$ [which is just (the degree-two homogeneous) $E_x^1[\rho]$, $E_x^1[\rho] = -J[\rho]$]; just as the Weizsäcker functional is the simplest expression for a kinetic energy density functional satisfying the two homogeneity scaling-conditions for $T_s[\rho]$ [14] (see also [40]). That $-J[\rho]$ is the simplest exchange energy functional satisfying the homogeneity conditions is of course stated from a physical point of view: a "real" exchange-energy density functional, on one hand, has to be a two-particle functional (regarding its structure), and on the other hand, a term $\frac{1}{|\vec{r}-\vec{r}\,'|}$, characterizing the interaction, has to appear explicitly beside the density-functional part, since the exchange energy is obtained via Eq.(10), the orbitals $u_i(\vec{r})$ forming the $\rho$-dependent part in the functional; thus, it has to have a form (or a combination of forms like)

$$\iint \frac{f[\rho(\vec{r})]f[\rho(\vec{r}\,')]}{|\vec{r}-\vec{r}\,'|} d\vec{r}\,d\vec{r}\,' \ . \tag{56}$$

Note that with the separation of a factor $\frac{\rho(\vec{r})\rho(\vec{r}\,')}{|\vec{r}-\vec{r}\,'|}$ in the integrand,

$$\iint g[\rho(\vec{r})]g[\rho(\vec{r}\,')]\frac{\rho(\vec{r})\rho(\vec{r}\,')}{|\vec{r}-\vec{r}\,'|} d\vec{r}\,d\vec{r}\,' \ , \tag{57}$$

the remaining part $g[\rho(\vec{r})]g[\rho(\vec{r}\,')]$ (or, simply $g[\rho(\vec{r})]$) has to be homogeneous of degree zero in $\rho(\vec{r})$ for Eq.(57) to satisfy the scaling conditions.

Finally, the question of separability [41] for $N$-particle density functionals may be worth taking a look at, referring also to a recent work [42], which investigates the separability problem of (explicitly) $N$-dependent density functionals. It can be seen easily that $N$-particle density functionals $A_N[\rho]$ corresponding to a separable density functional $A[\rho]$ are separable for $N$-particle densities (only for which they have physical relevance):

$$A_N[\rho_{N_1}+\rho_{N_2}] = A[\rho_{N_1}+\rho_{N_2}] = A[\rho_{N_1}]+A[\rho_{N_2}] = A_{N_1}[\rho_{N_1}]+A_{N_2}[\rho_{N_2}] \ , \tag{58}$$

where $N_1+N_2 = N$.

## IV. Summary

Various treatments of exchange in the framework of density-functional theory have been investigated. First, the so-called Hartree-Fock-Kohn-Sham method, which is the density-functional amplification of the Hartree-Fock approach, has been considered. In Sec.II, a simple derivation of the Hartree-Fock-Kohn-Sham equations has been presented, following the spirit of [14], showing how these equations, together with the Kohn-Sham equations and the ground-state Schrödinger equation [14], formally originate from one common root: density-functional theory, through its Hohenberg-Kohn Euler equation. Beside the Hartree-Fock-Kohn-Sham scheme, along the same lines, in Sec.II another mixture of density-functional theory and Hartree-Fock theory, the Kohn-Sham formulation of the Hartree-Fock approach (with the Kohn-Sham-Hartree-Fock equations), has also been considered.

In Sec.III, it has been pointed out that the exchange energy of ground-state systems of $N$ identical fermions naturally emerges as a second-degree homogeneous $N$-particle density functional $E_x^N[\rho]$, for any particle number $N$, yielding a construction $E_x[\rho] = E_x^{\int \rho}[\rho]$ of the exchange-energy density functional, for which Eq.(53) holds. The first element of the arising sequence (that is, $N=1,2,...$) of degree-two homogeneous $N$-particle exchange-energy density functionals $E_x^N[\rho]$ is just minus the classical Coulomb-repulsion energy functional. It has further been shown that $E_x^N[\rho]$ (homogeneous in $\rho$) retains the coordinate scaling property of the general exchange-energy density functional, $E_x[\rho]$, and generally, any $N$-particle density functional homogeneous in the density has the same coordinate scaling behaviour as the general density functional to which it corresponds. Thus, the scaling properties of $E_x^N[\rho]$ proposed in Sec.III are just the reverse of those of $T_s^N[\rho]$ of [14], which is first-degree homogeneous in $\rho$ and second-degree homogeneous with respect to coordinate scaling.

**Appendix: Uniqueness of homogeneous *N*-particle density functionals, and their coordinate scaling**

Given a density functional $A[\rho]$, $N$-particle density functionals $A_N[\rho]$ can be defined, for which

$$A_N[\rho_N] = A[\rho_N] \tag{A1}$$

(with $\int \rho_N(\vec{r})d\vec{r} = N$), giving the value of $A$ for $N$-particle densities $\rho_N$. The definition Eq.(A1) is of course ambiguous. E.g., the free choice of generalization of normalization represents the ambiguity in the case of $N$-particle noninteracting kinetic-energy density functionals constructed as described in [14]. Requiring an $A_N[\rho]$ to be homogeneous of some degree $k$, which means

$$A_N[\lambda\rho] = \lambda^k A_N[\rho] \,, \tag{A2}$$

however, leads to a unique definition for this $A_N[\rho]$. For, utilizing Eqs.(A1) and (A2),

$$\left(\frac{N}{\int\rho}\right)^k A_N[\rho] = A_N\left[N\frac{\rho}{\int\rho}\right] = A\left[N\frac{\rho}{\int\rho}\right] \,, \tag{A3}$$

making use of that the norm of $N\dfrac{\rho}{\int\rho}$ is $N$ [15]. From Eq.(A3) then

$$A_N[\rho] = \left(\frac{\int\rho}{N}\right)^k A\left[N\frac{\rho}{\int\rho}\right] \,, \tag{A4}$$

which shows that the fixation of $N$ (on the right) leads to the $N$-particle nature.

Eq.(A4) is the unique expression for a degree-$k$ homogeneous $N$-particle density functional corresponding to an $A[\rho]$. This uniqueness, in the case of the density functionals $T_s[\rho]$ and $E_x[\rho]$, means that the homogeneous $N$-particle density functionals $T_s^N[\rho]$ and $E_x^N[\rho]$ constructed in [14] and Sec.III of the present article are the only first-degree homogeneous $N$-particle noninteracting kinetic-energy and second-degree homogeneous $N$-particle exchange-energy density functionals, respectively. Thus, e.g., $T_W[\rho]$ and $-J[\rho]$ are the only possible first-degree homogeneous and second-degree homogeneous density functionals that give the exact noninteracting kinetic energy and the exact exchange energy, respectively, for one-electron systems.

With Eq.(A4), the validity of Eq.(54) [and Eq.(55)] for arbitrary $\rho(\vec{r})$ (not just for $\rho(\vec{r})$'s of norm $N$), i.e.

$$E_x^N[\rho] = -\int \rho(\vec{r})\vec{r}\nabla\frac{\delta E_x^N[\rho]}{\delta\rho(\vec{r})}d\vec{r} \,, \tag{A5}$$

can be justified easily. Utilizing $\int \lambda^3 \rho(\lambda \bar{r}) d\bar{r} = \int \rho(\bar{r}) d\bar{r}$ and the homogeneity of degree one of $E_x[\rho]$ in coordinate scaling,

$$E_x^N[\lambda^3 \rho(\lambda \bar{r})] = \left(\frac{\int \rho}{N}\right)^2 E_x\left[\lambda^3 \frac{N}{\int \rho} \rho(\lambda \bar{r})\right] = \left(\frac{\int \rho}{N}\right)^2 \lambda E_x\left[\frac{N}{\int \rho} \rho(\bar{r})\right] = \lambda E_x^N[\rho(\bar{r})], \quad (A6)$$

which means that $E_x^N[\rho]$ itself, too, is homogeneous of degree one in coordinate scaling, yielding Eq.(A5). A similar proof applies for $T_s^N[\rho]$ of [14] as well, meaning that $T_s^N[\rho]$ has the same coordinate scaling property as $T_s[\rho]$, that is, homogeneity of degree two.

**Acknowledgements:** Grants from the Hungarian Higher Education And Research Foundation and from OTKA (D048675) are gratefully acknowledged.

**References**

[1] Parr, R. G., and Yang, W., *Density Functional Theory of Atoms and Molecules* (Oxford Univ. Press, New York, 1989).

[2] Nagy, Á., Phys. Rep. **298**, 1 (1998).

[3] Thomas, L. H., Proc. Cambridge Phil. Soc. **23**, 542 (1927).

[4] Fermi, E., Z. Phys. **48**, 73 (1928).

[5] Dirac, P. A. M., Proc. Cambridge. Phil. Soc. **26**, 376 (1930).

[6] von Weizsäcker, C. F., Z. Phys. **96**, 431 (1935).

[7] March, N. H., Adv. Phys. **6**, 1 (1957).

[8] Slater, J. C., Phys. Rev. **81**, 385 (1951).

[9] Gáspár, R., Acta Phys. Hung. **3**, 263 (1954) [for English translation, see J. Mol. Struct. Theochem **501**, 1 (2000)].

[10] Slater, J. C., *The Self-Consistent Field for Molecules and Solids, Quantum Theory of Molecules and Solids, Vol. 4* (McGraw-Hill, New York, 1974).

[11] Hohenberg, P., and Kohn, W., Phys. Rev. **136**, B864 (1964).

[12] Kohn, W., and Sham, L. J., Phys. Rev. **140**, A1133 (1965).

[13] Holas, A., and March, N. H., in *Density Functional Theory*, Topics in Current

Chemistry, Vol. 180, ed. R. F. Nalewajski (Springer, Heidelberg, 1996), p. 57.

[14] Gál, T., Phys. Rev. A **64**, 062503 (2001).

[15] Gál, T., Int. J. Quantum Chem. **107**, 2586 (2007).

[16] Percus, J. K., Int. J. Quantum Chem. **13**, 89 (1978).

[17] Levy, M., Proc. Natl. Acad. Sci. USA **76**, 6062 (1979).

[18] Lieb, E. H., Int. J. Quantum Chem. **24**, 243 (1983).

[19] Payne, P. W., J. Chem. Phys. **71**, 490 (1979).

[20] Freed, K. F., and Levy, M., J. Chem. Phys. **77**, 396 (1982).

[21] Holas, A., March, N. H., Takahashi, Y., Zhang, C., Phys. Rev. A **48**, 2708 (1993).

[22] Gál, T., Phys. Rev. A **63**, 022506 (2001).

[23] Gál, T., J. Math. Chem. **42**, 661 (2007) [eprint arXiv:math-ph/0603027].

[24] Perdew, J.P., Parr, R.G., Levy, M., Balduz, J.L., Phys. Rev. Lett. **49**, 1691 (1982).

[25] Levy, M., and Perdew, J. P., Phys. Rev. A **32**, 2010 (1985).

[26] Appendix in Ghosh, S. K., and Parr, R. G., J. Chem. Phys. **82**, 3307 (1985).

[27] Zhao, Q., Morrison, R. C., and Parr, R. G., Phys. Rev. A **50**, 2138 (1994).

[28] Parr, R. G., and Ghosh, S. K., Phys. Rev. A **51**, 3564 (1995).

[29] Liu, S., and Parr, R. G., Phys. Rev. A **53**, 2211 (1996).

[30] Morrison, R. C., and Parr, R. G., Phys. Rev. A **53**, R2918 (1996).

[31] Liu, S., and Parr, R. G., Phys. Rev. A **55**, 1792 (1997).

[32] Parr, R. G., and Liu, S., Chem. Phys. Lett. **276**, 164 (1997).

[33] Liu, S., and Parr, R. G., Chem. Phys. Lett. **278**, 341 (1997).

[34] Parr, R. G., and Liu, S., Chem. Phys. Lett. **280**, 159 (1997).

[35] Joubert, D. P., Chem. Phys. Lett. **288**, 338 (1998).

[36] Tozer, D. J., Phys. Rev. A **58**, 3524 (1998).

[37] Chan, G. K.-L., and Handy, N. C., Phys. Rev. A **59**, 2670 (1999).

[38] Gál, T., Phys. Rev. A **62**, 044501 (2000).

[39] Joubert, D., Phys. Rev. A **64**, 054501 (2001).

[40] Gál, T., and Nagy, Á., J. Mol. Struct. Theochem **501**, 167 (2000).

[41] Perdew, J. P., Adv. Quantum Chem. **21**, 113 (1990).

[42] Pérez-Jiménez, A. J., Moscardó, F., Sancho-García, J. C., Abia, L. P., San-Fabián, E., and Pérez-Jordá, J. M., J. Chem. Phys. **114**, 2022 (2001).

**Appendix B.** Summary of the connections between the density-functional Euler equation and the various wave-function equations, indicating the functional the constrained-search definition of which is utilized to get to the given wave-function equation(s)

*Exact quantum mechanics*:

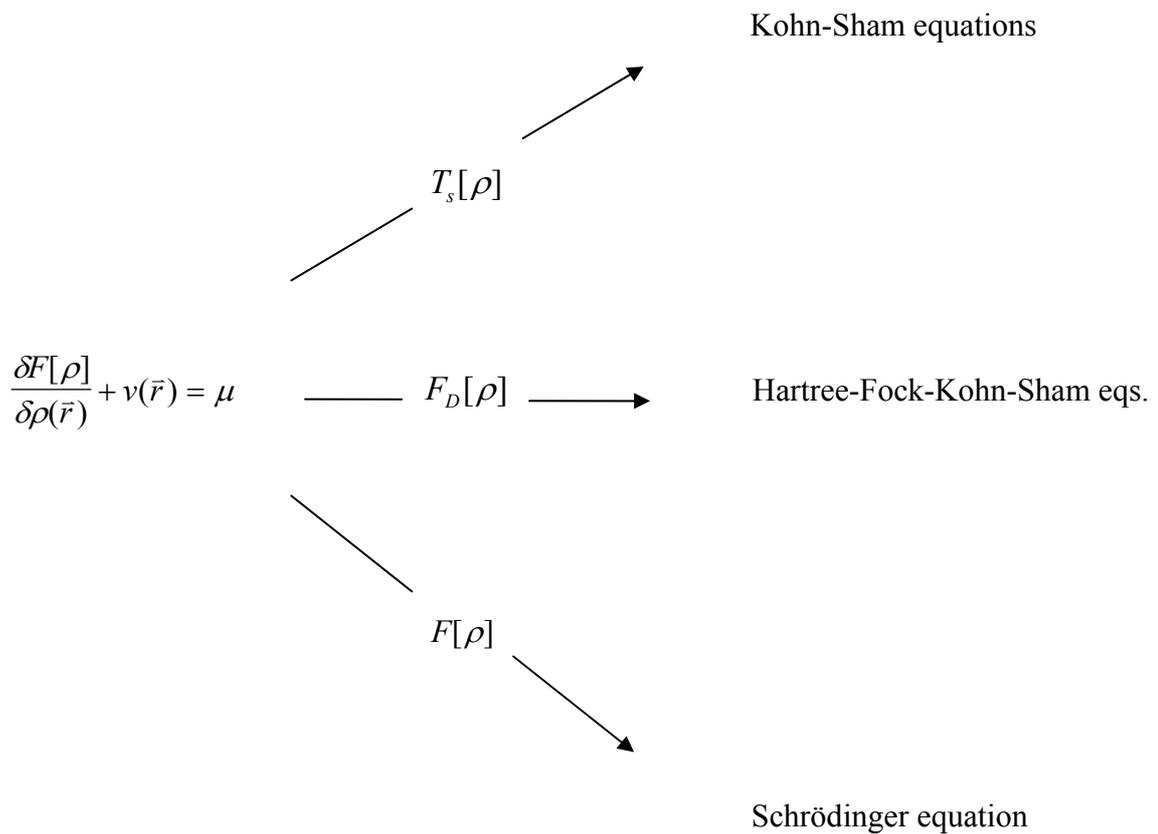

*Hartree-Fock approximation*:

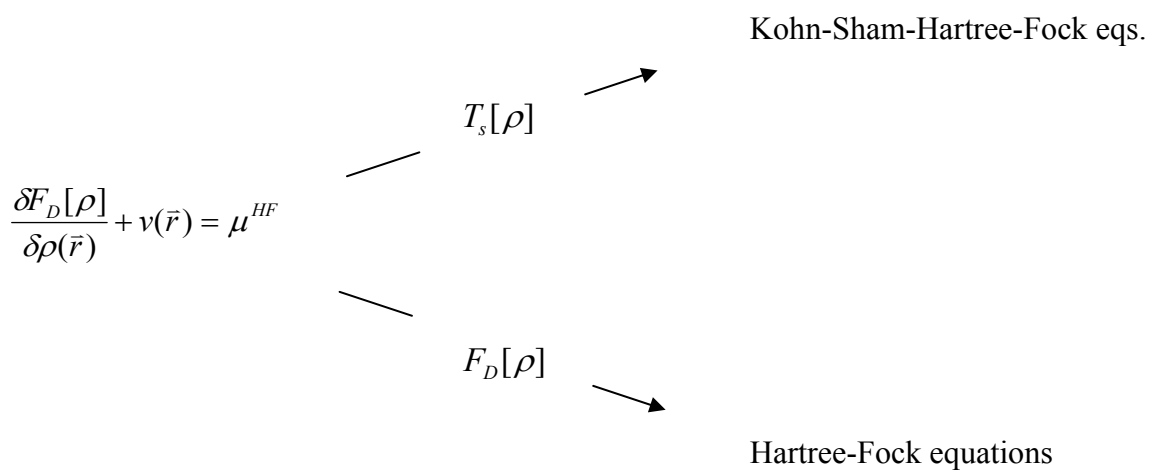